\documentclass[aps,prd,twocolumn,groupedaddress]{revtex4-1}
\usepackage{graphicx}
\usepackage{epsfig}
\usepackage{amsmath}
\usepackage{amssymb}
\usepackage{color}

\newcommand{\stau}{\sqrt{\tau}}

\begin{document}

\title{Geometrical Scaling \\in Inelastic Inclusive Particle Production
at the LHC }
\author{Micha\l \ Prasza\l owicz and Anna Francuz}
\affiliation{M. Smoluchowski Institute of Physics, Jagiellonian University, 
{\L}ojasiewicza 11, 30-348 Krak{\'o}w, Poland.}

\date{\today}

\begin{abstract}
Analyzing recent ALICE data on inelastic pp scattering at the LHC energies we show
that charged particle distributions exhibit geometrical scaling (GS). 
We show  also that the
inelastic cross-section is  scaling as well and that in this case the quality of  GS is
better than for multiplicities. Moreover, exponent $\lambda$ characterizing the
saturation scale is for the cross-section scaling compatible with the one found in
deep inelastic ep scattering at HERA. Next, by parametrizing charged particles
distributions by the Tsallis-like formula, we find a somewhat unexpected solution
that still exhibits GS, but differs from the "standard" one where the Tsallis 
temperature is proportional to the saturation scale.

\end{abstract}

\maketitle

\section{Introduction}

It is believed that gluon distribution inside a hadron saturates at small Bjorken $x$
(see Refs.~\cite{Mueller:2001fv,McLerran:2010ub} for review).
This is a consequence of the non-linear QCD evolution equations, known as Balitsky-Kovchegov
(BK) equation \cite{BK} or in a more general case as JIMWLK equation \cite{jimwlk}, that poses 
traveling wave solutions \cite{Munier:2003vc}. The latter property, in the QCD context, 
is called geometrical
scaling (GS) \cite{Stasto:2000er}. 
An
effective theory relevant for the small Bjorken $x$ region is so called Color
Glass Condensate (CGC) \cite{MLV}. For the purpose of present work the details
of the saturation are not of primary importance; it is the very existence of the
saturation scale which plays the crucial role

Geometrical scaling  means that some observable $N$ that in
principle depends on two independent kinematical variables, say $x$ and
$Q^{2}$, in fact depends only on a specific combination of them denoted as
$\tau$:
\begin{equation}
N(x,Q^{2})= {\cal F}(\tau). \label{GSdef}%
\end{equation}
Here function ${\cal F}$ in Eq.~(\ref{GSdef}) is a dimensionless function of a scaling
variable \cite{fn1}
\begin{equation}
\tau=Q^{2}/Q_{\rm  s}^{2}(x). \label{taudef}%
\end{equation}
and
\begin{equation}
Q_{\rm  {s}}^{2}(x)=Q_{0}^{2}\left(  {x}/{x_{0}}\right)  ^{-\lambda}
\label{Qsat}%
\end{equation}
is the saturation scale. The power law form of the saturation scale is dictated
by a saddle-point solution to the BK equation and has been tested phenomenologically
for different high energy processes 
\cite{Stasto:2000er}\nocite{McLerran:2010ex,Praszalowicz:2012zh}--\cite{Praszalowicz:2013uu}.

Here we are coming back to pp scattering \cite{McLerran:2010ex}
in the context of recently published ALICE
data \cite{Abelev:2013ala} for charged particle distributions at three LHC energies 0.9, 2.76 and 7 TeV.
After discussing shortly in Sect.~\ref{sec:GSHC} how GS emerges in the 
$k_{\rm   T}$ factorization scheme, we shall show in Sect.~\ref{sec:GSALICE}
that recent ALICE data indeed exhibit geometrical scaling with, however, $\lambda$
exponent being different than in the case of deep inelastic (DIS) ep scattering. Interestnigly,
we shall also show that the inclusive cross-sections scale somewhat better and with 
an exponent that is very close to the DIS value $\lambda=0.32$ \cite{Praszalowicz:2012zh}. 
This result calls for better understanding
of the impact parameter picture of pp scattering in the context of the saturation physics and 
the Color Glass Condensate theory. 

Another topic addressed in the present paper is the shape of the scaling function
introduced schematically in Eq.~(\ref{GSdef}). Function $\cal{F}$ can be in fact obtained
only numerically within some specific model. Here, we shall use phenomenological
parametrization in the form of Tsallis-like  distribution \cite{Tsallis} applied successfully in the past
to describe spectra of charged particles \cite{Wong:2012zr}\nocite{Cirto:2014sra}--\cite{Cleymans:2013rfq}. 
In Sect.\ref{sec:GSTsallis} we briefly
describe how GS should be reflected in the Tsallis distribution. Next, 
in Sect.~\ref{sec:TsALICE} we shall
try do fit Tsallis-like parametrization to the ALICE data. Unfortunately,
as already remarked in the original ALICE publication \cite{Abelev:2013ala}, this piece of data does
not admit good quality Tsallis fit. Nevertheless, we  invoke
a procedure that allows for rather good description of the data in the range of moderate
transverse momenta where GS is expected to occur. Somewhat unexpectedly
we find GS scaling solution that is very different from the "standard" one
described in Sect.~\ref{sec:GSTsallis}. Unfortunately GS in this solution is rather 
accidental and will disappear at very high energies. Whether this is only a
property of the Tsallis parametrization "forced" to describe ALICE data, or
a real prediction, remains to bee seen. We conclude in Sect.~\ref{sec:Concl}

\section{Geometrical scaling in hadronic collisions}
\label{sec:GSHC}

The cross-section for producing a moderate $p_{\rm  {T}}$ gluon in hadronic
collision can be described as \cite{Gribov:1981kg}:%
\begin{equation}
\frac{d\sigma}{dyd^{2}p_{\rm  T}}=\frac{C}{p_{\rm   {T}}^{2}}%
{\displaystyle\int}
d^{2}\vec{k}_{\rm    {T}}\,\alpha_{\rm    {s}}(k_{\rm    {T}}^{2})\varphi_{1}%
(x_{1},\vec{k}_{\rm  {T}}^{2})\varphi_{2}(x_{2},(\vec{k}-\vec{p})_{\rm  {T}%
}^{2})\label{Nchdef}%
\end{equation}
where  $C$ contains color factors and numerical constants. Bjorken $x$'s of
colliding partons read%
\begin{equation}
x_{1,2}=\frac{p_{\mathrm{T}}}{\sqrt{s}}\,e^{\pm y}%
\end{equation}
however, since in the following we will be interested in central rapidity
production, \emph{i.e.} $y\simeq0$, we have $x_{1}\simeq x_{2}=x$. Functions
$\varphi_{1,2}$ are the unintegrated gluon distributions in hadron $h_{1}$ and
$h_{2}$ respectively. There exist many models of unintegrated gluon
distributions (see {\em e.g.} \cite{Szczurek:2003vq}); the most simple ones are the one of the
Golec-Biernat--W{\"u}sthoff (GBW) model ~\cite{GolecBiernat:1998js} 
or the one by Kharzeev, Levin and Nardi~\cite{Kharzeev:2002ei}. They
share two important features: geometrical scaling and dependence on the
transverse area $S_{\bot}$ whose precise meaning is best understood in a picture where
also the impact parameter is taken into account~\cite{Kowalski:2003hm,Tribedy:2010ab}. 
Therfore%
\begin{equation}
\varphi(x,\vec{k}_{\rm  {T}}^{2})=S_{\bot}\phi(k_{\rm  {T}}^{2}%
/Q_{\rm  {s}}^{2}(x))
\end{equation}
where $\phi$ is a dimensionless function of the scaling variable $k_{\rm  {T}%
}^{2}/Q_{\rm  {s}}^{2}(x)$, rather than independently of $x$ and $k_{\rm  {T}%
}^{2}$. Of course geometrical scaling is only an approximation and is expected
to break for large Bjorken $x$'s and also for large transverse momenta. We also
expect GS breaking for small $k_{\rm   T}$ where non-perturbative effects including
effects from the pion mass are  of importance.
Ignoring these effects and neglecting also momentum dependence of the strong coupling
constant we arrive at%
\begin{equation}
\frac{d\sigma}{dyd^{2}p_{\rm  {T}}}=\frac{S_{\bot}^{2}}{2\pi}\mathcal{F}%
(\tau)\label{sigmaGS}%
\end{equation}
where $\mathcal{F}$ is a universal, energy independent function of the scaling variable
$\tau$:
\begin{equation}
\tau=\frac{p_{\rm  {T}}^{2}}{Q_{\rm  {s}}^{2}(x)}=\left(  \frac{p_{\rm  {T}}%
}{Q_{0}}\right)  ^{2}\left(  \frac{p_{\rm  {T}}}{\xi W}\right)  ^{\lambda
}.\label{tau}%
\end{equation}
Here we have used (\ref{Qsat}) for the saturation scale $Q_{\mathrm{s}%
}^{2}$.  We take for $x_{0}=\xi\times10^{-3}.$
This implies that in (\ref{tau}) $p_{\rm  {T}}$ is in GeV$/c$ and $W$ in TeV.
Furthermore for $Q_{0}$ we can take without any loss of generality $Q_{0}=1$
GeV$/c.$ One typically assumes that $S_{\bot}$ is an energy independent
constant. This is true in the case of the GBW model \cite{GolecBiernat:1998js}
where $S_{\bot}=\sigma
_{0}$ with $\sigma_{0}$ characterizing  the asymptotics of the dipole-proton 
cross-section for large dipole sizes. In the
case of heavy ion collisions for fixed centrality, $S_{\bot}$ has geometrical
interpretation as an overlap area of two colliding nuclei~\cite{Kharzeev:2002ei}. 
In this case one
can also assume that%
\begin{equation}
\frac{d\sigma}{dyd^{2}p_{\rm  {T}}}=S_{\bot}\frac{d^{2}N}{dyd^2 p_{\rm  {T}}%
}\label{sigma_vs_N}%
\end{equation}
where $N$ is a multiplicity of produced gluons. Neglecting possible energy
dependence of gluon fragmentation into hadrons \cite{Levin:2011hr}, \emph{i.e.}
adopting parton-hadron duality hypothesis \cite{PHD}, we arrive at:%
\begin{equation}
\frac{1}{p_{\rm  {T}}}\frac{d^{2}N_{\rm  {ch}}}{dydp_{\rm  {T}}}=S_{\bot
}\mathcal{F}(\tau).\label{dNdydef}%
\end{equation}
Expression (\ref{dNdydef})  will be used in the following to look for GS in
the multiplicity distributions. Let us, however, note that GS is in fact a
property of  Eq.(\ref{sigmaGS}) and that the multiplicity scaling
(\ref{dNdydef}) is based on (\ref{sigma_vs_N}) which is not so obvious for the
scattering of small systems, like pp.

In order to integrate (\ref{dNdydef}) over $d^{2}p_{\rm  {T}}$ we have to
change variables%
\begin{equation}
p_{\rm  {T}}=\overline{Q}_{\rm  {s}}(W)\tau^{1/(2+\lambda)} \label{pT}%
\end{equation}
where the \emph{average} saturation scale is defined as%
\begin{equation}
\overline{Q}_{\rm  {s}}(W)=Q_{0}\left(  \frac{\xi W}{Q_{0}}\right)
^{\lambda/(2+\lambda)}. \label{Qbarsat}%
\end{equation}
Note that the effective power describing the rise of the average saturation
scale with energy $\lambda_{\mathrm{eff}}=\lambda/(2+\lambda)$ is slightly
smaller than $\lambda/2$. Then%
\begin{equation}
p_{\rm  {T}}dp_{\rm  {T}}=\frac{1}{2+\lambda}\overline{Q}_{\rm  {s}}%
^{2}(W)\,\tau^{2/(2+\lambda)}\frac{d\tau}{\tau}.
\end{equation}
Hence%
\begin{eqnarray}
\frac{dN_{\rm   ch}}{dy}& = & \left[  \frac{1}{2+\lambda}%
{\displaystyle\int}
\mathcal{F}(\tau)\tau^{2/(2+\lambda)}\frac{d\tau}{\tau}\right]  \times
S_{\bot} \overline{Q}_{\rm  {s}}^{2}(W) \nonumber \\
&=& b \times
S_{\bot} \overline{Q}_{\rm  {s}}^{2}(W) \ .
\label{dNdy}
\end{eqnarray}
Here $b$ is an energy independent constant related to the the integral of
$\mathcal{F}(\tau)$. Equation (\ref{dNdy}) is often used as a definition of the
saturation scale (with $N_{\rm   ch}$ replaced by $N_{\rm   gluons}$) understood
as gluon number density per transverse area.

\section{Geometrical scaling of ALICE data}

\label{sec:GSALICE}

\begin{figure*}[t]
\centering
\includegraphics[width=7.0cm]{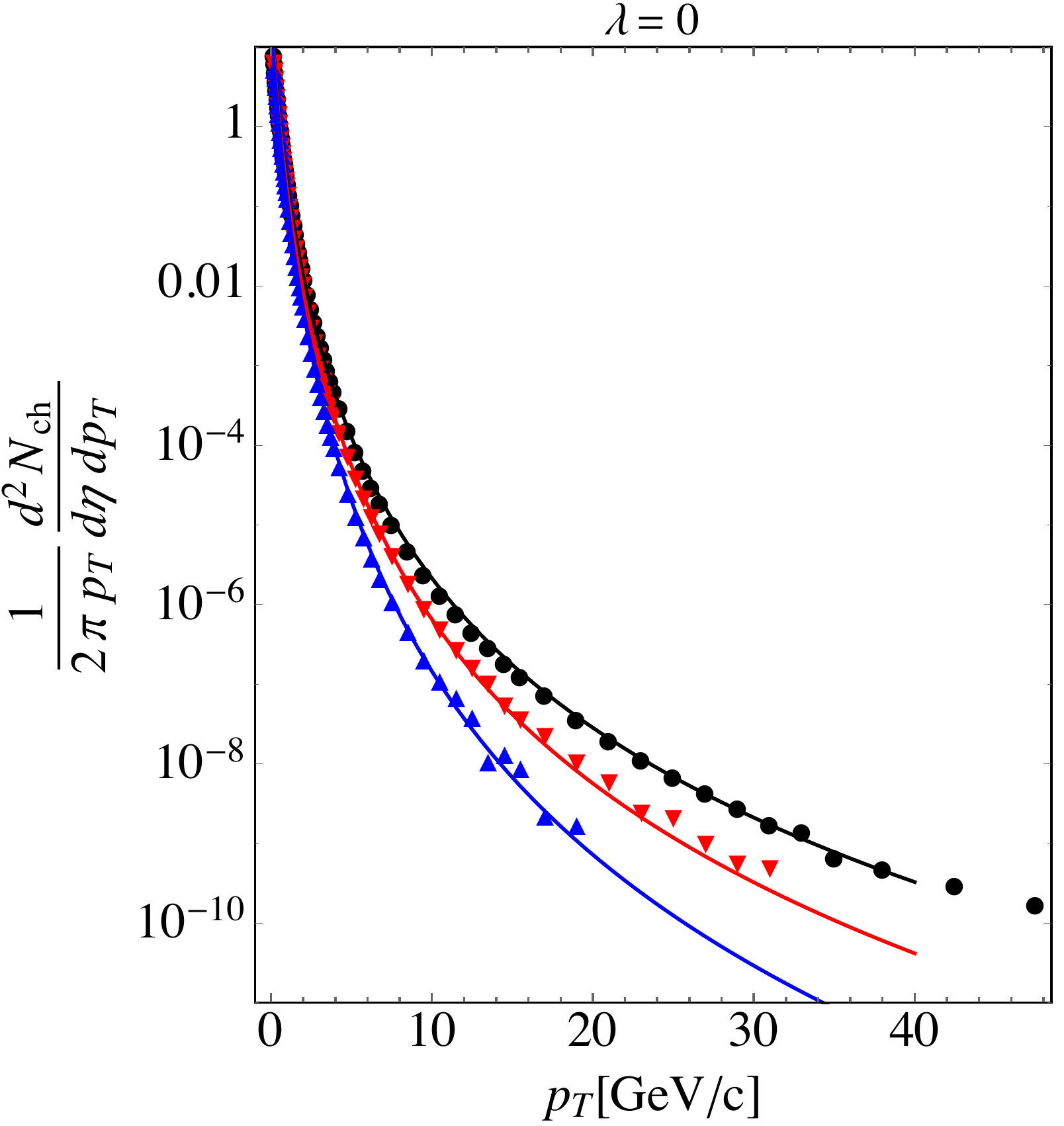}~~~
\includegraphics[width=7.0cm]{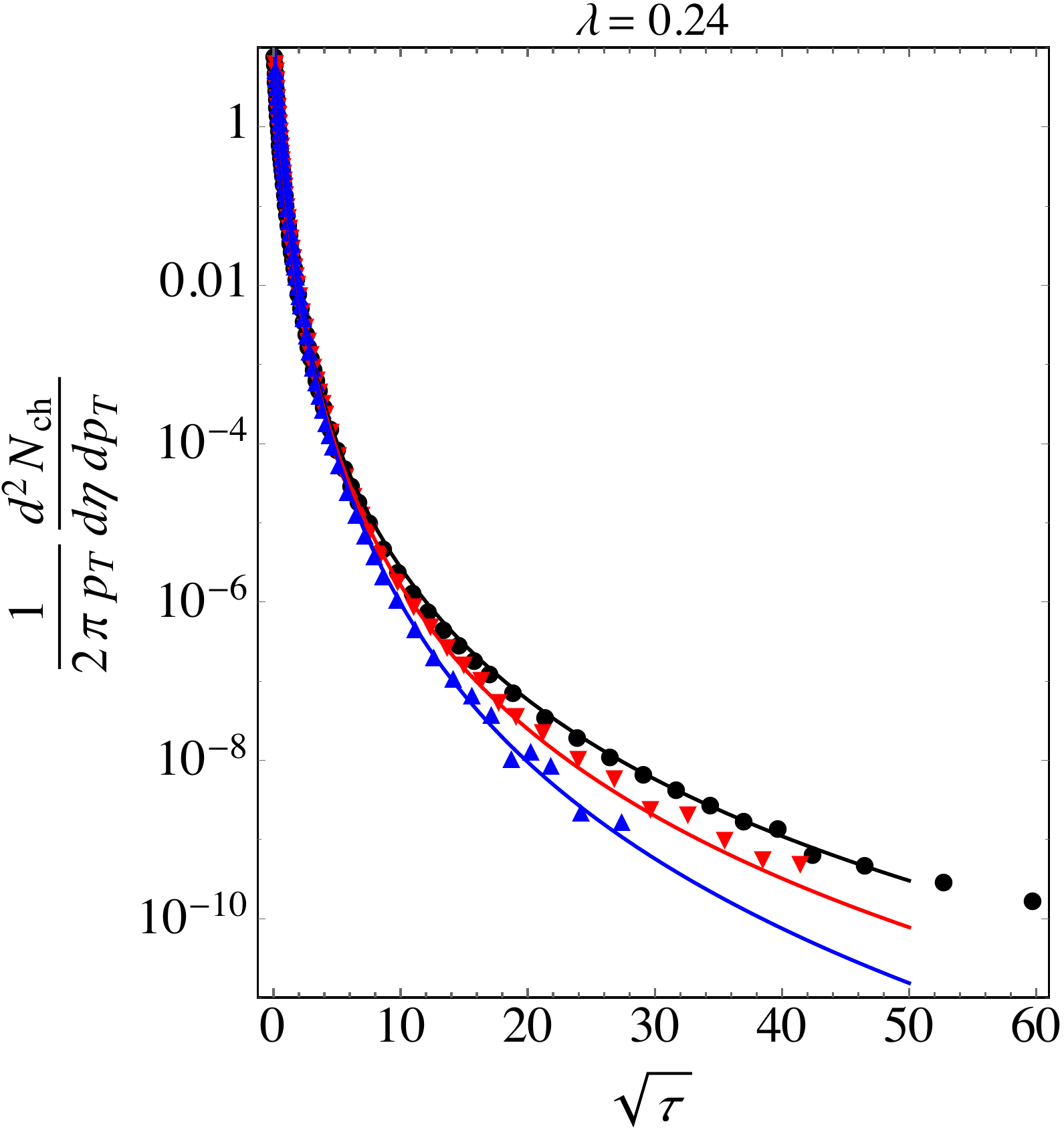} \caption{Charged
particle spectra measured by ALICE~\cite{Abelev:2013ala}, plotted as functions of $p_{\mathrm{T}}$
(left panel) and as functions of the scaling variable $\tau$ (\ref{tau}) for
$\lambda=0.24$ (right panel). Black full dots correspond to $W=7$~TeV, red
down-triangles to 2.76~TeV and blue up-triangles to 0.9~TeV. Solid lines
correspond to the Tsallis fits from Sect.~\ref{sec:TsALICE}.}%
\label{fig:spectra}%
\end{figure*}

In this Section we are going  to check whether ALICE data~\cite{Abelev:2013ala}
on inelastic multiplicity distributions of charged particles
exhibits geometrical scaling and for what value of $\lambda$. We shall show
that indeed GS is reached for $\lambda\sim0.22 - 0.24$ as it is illustrated in
Fig.~\ref{fig:spectra}.

\begin{figure*}[t]
\centering
\includegraphics[width=6.1cm]{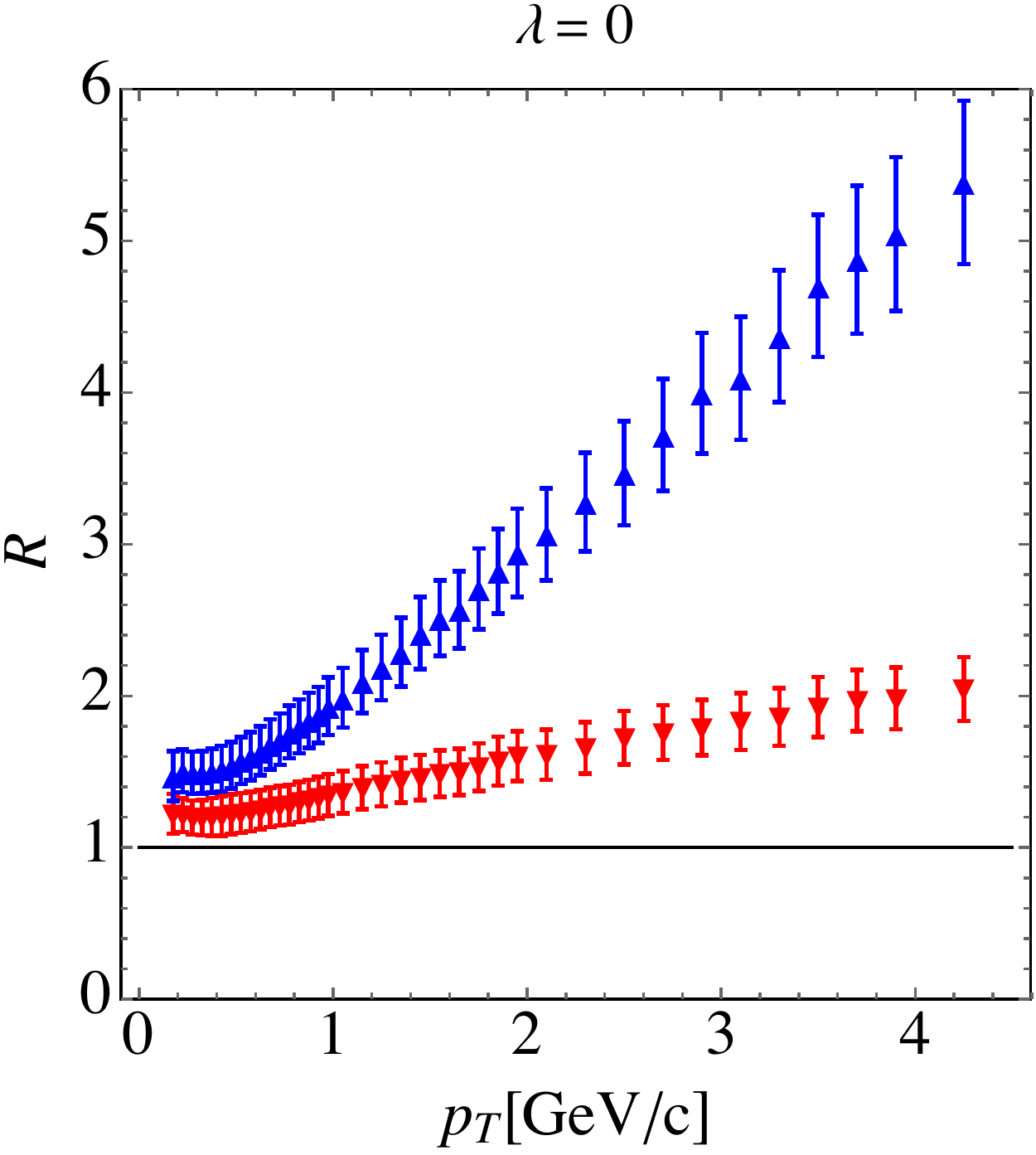}~~~
\includegraphics[width=6.0cm]{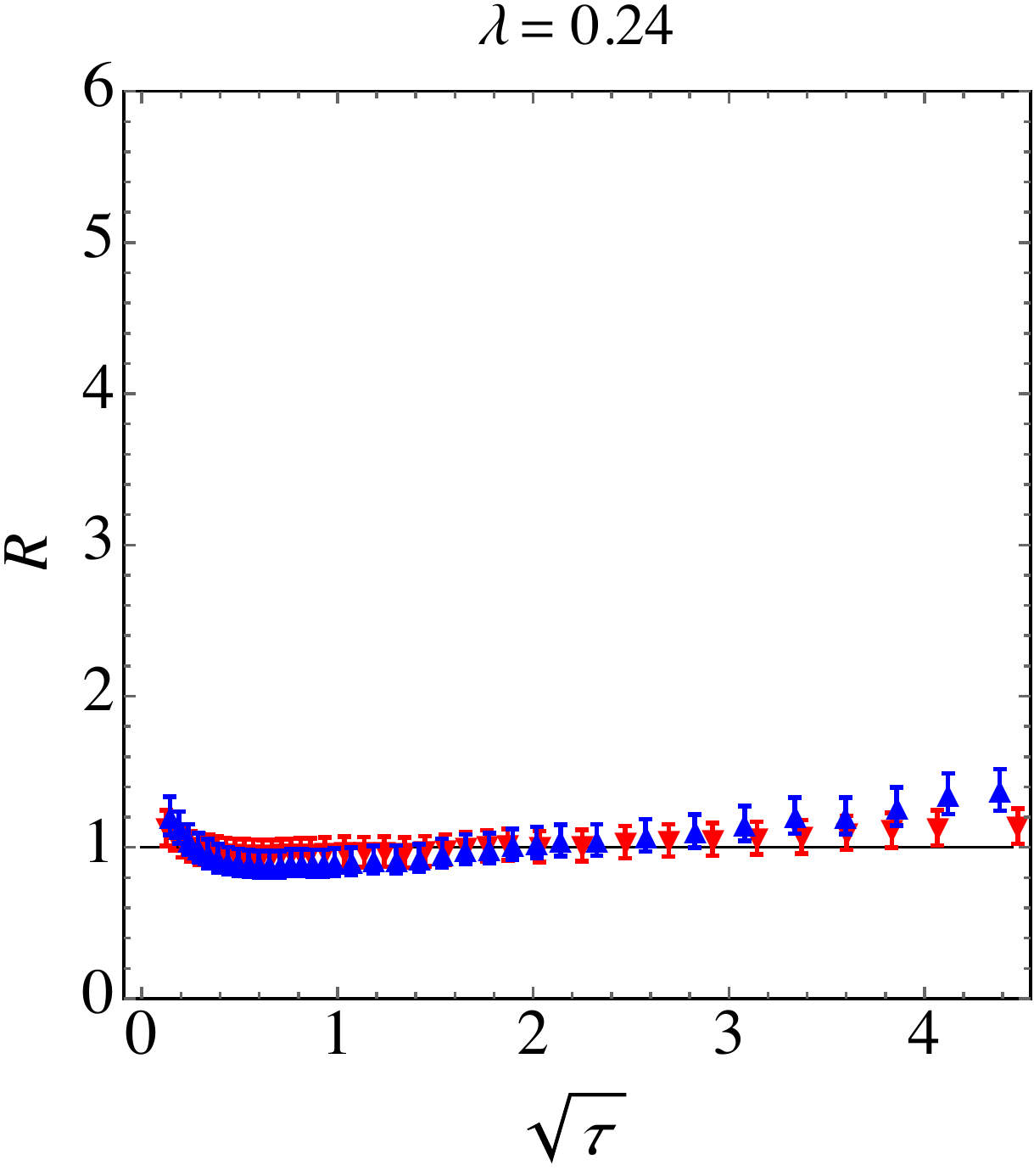}
\caption{Ratios of charged
particle spectra measured by ALICE~\cite{Abelev:2013ala}, plotted as functions of 
$p_{\mathrm{T}}$
(left panel) and as functions of the scaling variable $\tau$ (\ref{tau}) for
$\lambda=0.24$ (right panel). Red
down-triangles correspond to  to the ratio 7/2.76~TeV and blue up-triangles 
to 7/0.9~TeV. }%
\label{fig:ratios}%
\end{figure*}

In order to find the best value of $\lambda$ we have adopted the \emph{method
of ratios} described in more detail in Refs.~\cite{Praszalowicz:2012zh}. Let us denote for
simplicity%
\begin{equation}
N(p_{\rm  {T}},W)=\left.  \frac{1}{2\pi p_{\rm  {T}}}\frac{d^{2}%
N_{\rm   ch}}{dydp_{\rm  {T}}}\right\vert _{W}.
\end{equation}
We form ratios of spectra expressed in terms of the scaling variable $\tau$
rather than in terms of $p_{\mathrm{T}}$:
\begin{equation}
R_{W/W^{\prime}}(\tau)=\frac{N(\tau,W)}{N(\tau,W^{\prime})}%
\label{Rdef}
\end{equation}
and request that $R \sim1$ over the largest possible interval of $\tau$. Note
that this method is sensitive only to the value of $\lambda$ and not to 
the actual values of parameters $Q_{0}$ and
$x_{0}$
In the present case we choose 7 TeV for $W$ and $W_{1}
=2.76$ or $W_{2}=0.9$~TeV for $W^{\prime}$. Therefore we can form two such
ratios, which are depicted in Fig.~\ref{fig:ratios} for $\lambda=0$
(\emph{i.e.} for $\sqrt{\tau}=p_{\mathrm{T}}$) and for $\lambda=0.24$. We see
that indeed, these two ratios that rise rather steeply with $p_{\mathrm{T}}$,
remain flat and close to 1 if plotted in terms of $\sqrt{\tau}$ for
$\lambda=0.24$. We interpret this as a signature of geometrical scaling.

\begin{figure*}[t]
\centering
\includegraphics[width=6.0cm]{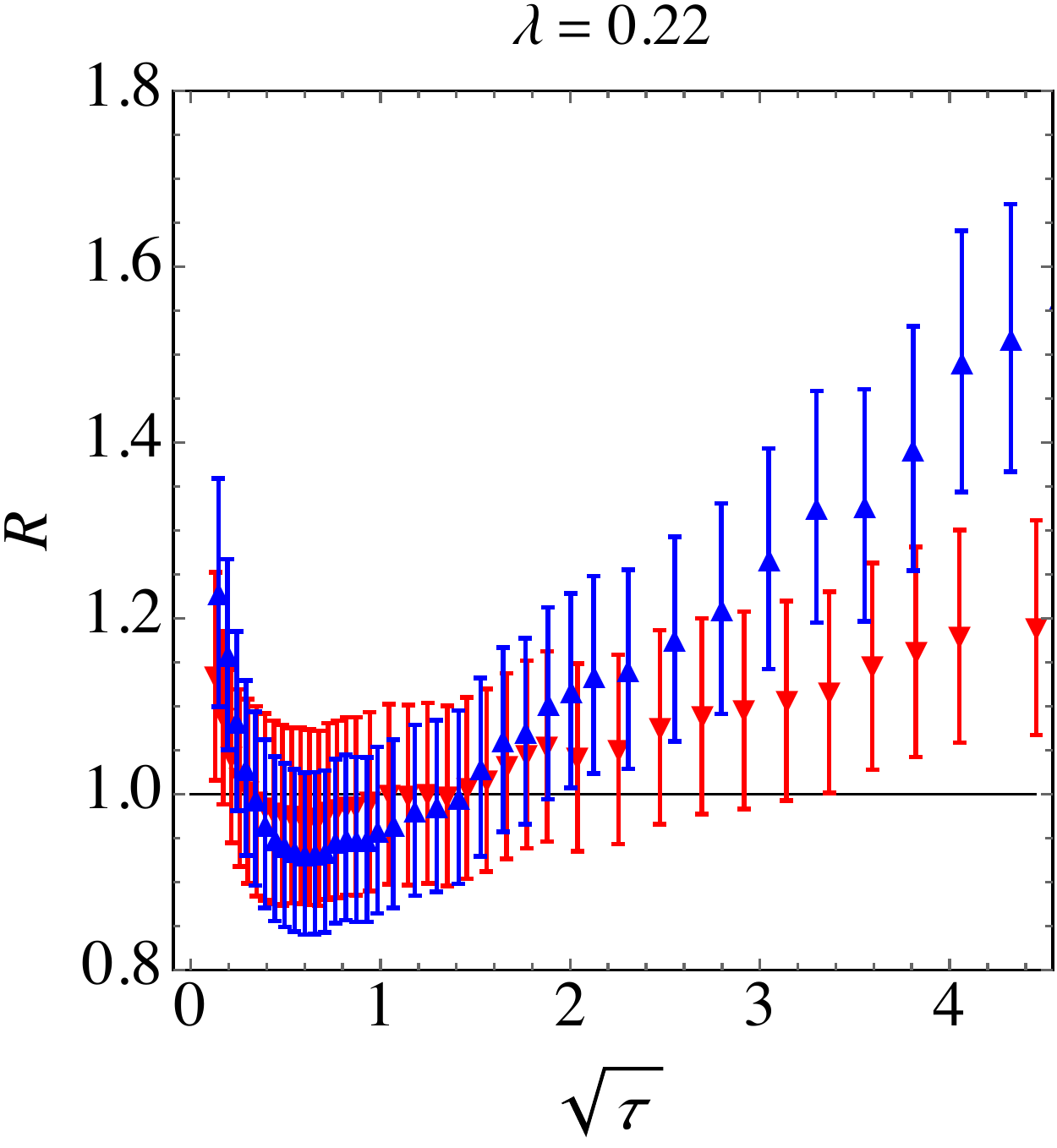}~~~
\includegraphics[width=6.0cm]{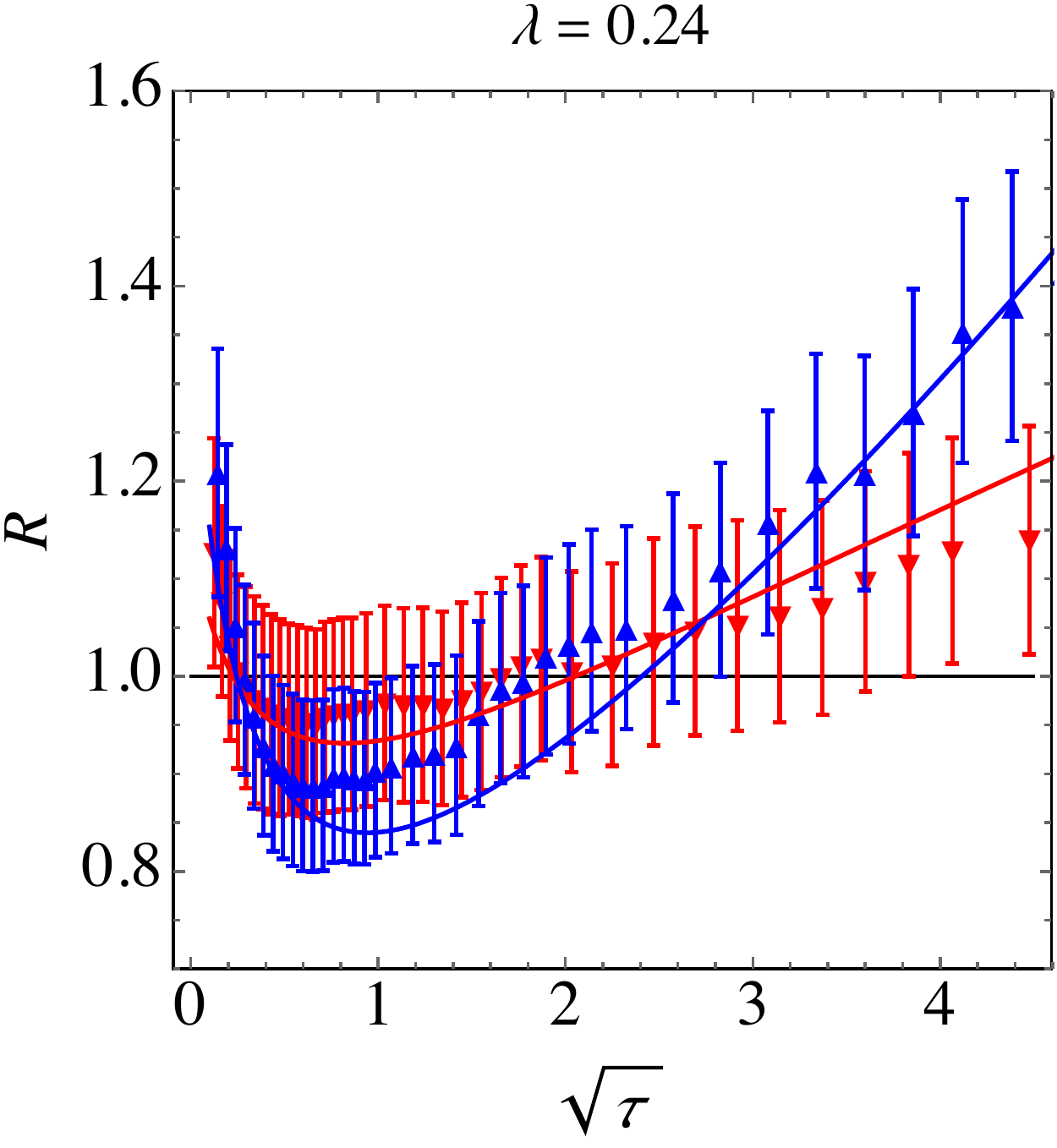}
\caption{Same as Fig.~\ref{fig:ratios} with different scale for better resolution.
Solid lines in the right panel correspond to the Tsallis fits from Sect.~\ref{sec:TsALICE}.}%
\label{fig:ratios2}%
\end{figure*}

In order to decide on the best value of exponent $\lambda$ we need to provide
a quantitative criterion measuring the "average distance" of experimental
values of $R_{W/W^{\prime}}$ from unity.
To this end we propose the following procedure. Since for  $\lambda$ values
relevant for the present analysis the first few $R$ points corresponding to
low $p_{\mathrm{T}}$ lie  above 1 (which is the sign of GS violation in a
region when  non-perturbative effects are of
importance) we pick up the first point for which
\[
R_{W/W_{1,2}}(\tau_{\mathrm{start}})-1\le\Delta_{R}(\tau_{\mathrm{start}}).
\]
Here $\Delta_{R}$ is an experimental error of $R$. For points with $\tau
>\tau_{\mathrm{start}}$ ratio $R$ is either close to 1 within the experimental
errors, or it is falling below 1 exceeding $\Delta_{R}$. Next, since for large
transverse momenta $p_{\mathrm{T}}$ spectra are getting harder with increasing
energy, the values of $R$ start to increase with $\tau$, getting again larger
than 1. This is well visible in Fig.~\ref{fig:ratios2} where the vertical
scale has been magnified with respect to Fig.~\ref{fig:ratios} for better resolution.

\begin{figure}[h]
\centering
\includegraphics[width=6.0cm]{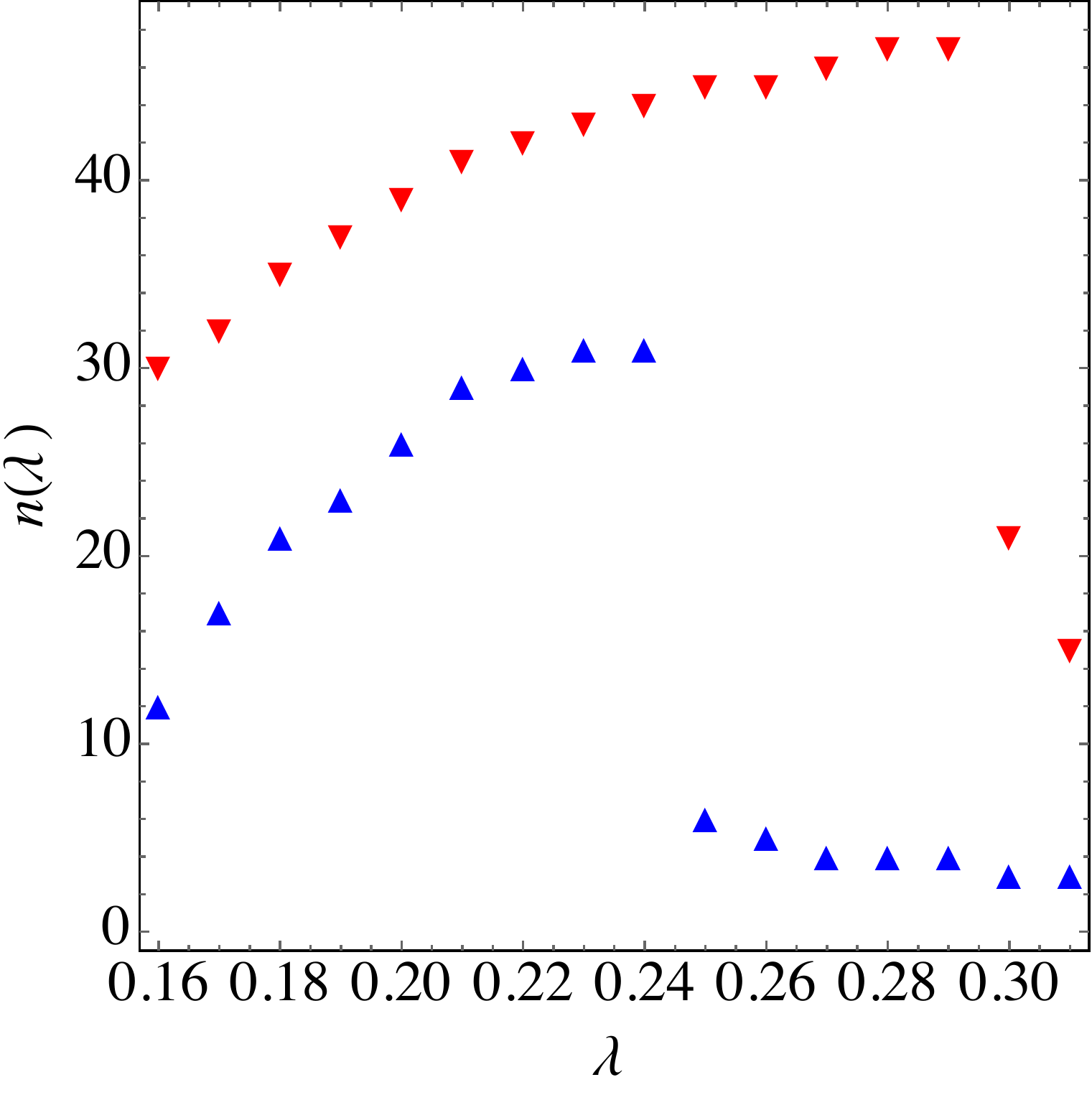} 
\caption{Number of points that contribute to (\ref{delta}) for two different ratios:
7/2.76~TeV (red down-triangles) and  7/0.9~TeV (blue up-triangles) plotted
as functions of $\lambda$.}%
\label{fig:np}%
\end{figure}

Starting from $\tau_{\mathrm{start}}$ that of course depends on energy
$W_{1,2}$, we compute mean square deviation for given $\lambda$
\begin{equation}
\delta_{W_{1,2}}^{2}(\lambda)=\frac{1}{n_{W_{1,2}}(\lambda)}%
{\displaystyle\sum\limits_{\tau_{n} =\tau_{\mathrm{satrt}}}^{\tau
_{\mathrm{end}}}}
\frac{\left(  R(\tau_{n},W_{1,2})-1\right)  ^{2}}{\Delta_{R} ^{2}(\tau
_{n},W_{1,2})} \label{delta}%
\end{equation}
where $n_{W_{1,2}}(\lambda)$ is a number of points between $\tau
_{\mathrm{start}}$ and $\tau_{\mathrm{end}}$. We increase $\tau_{\mathrm{end}%
}$ up to the last point where $\delta_{W_{1,2}}^{2}(\lambda)<1$. In this way
we obtain $n_{W^{\prime}}(\lambda)$ which is the number of points that exhibit
GS for given $W^{\prime}=W_{1,2}$ and for given $\lambda$, which are plotted
in Fig.~\ref{fig:np}.
Now we look for maximum
of $n_{W_{1}}(\lambda)+n_{W_{2}}(\lambda)$. This happens for $\lambda=0.24$.
As seen from Fig.~\ref{fig:ratios2} $W_{2}=0.9$~TeV points scale in a shorter
interval of $\tau$, which translated back to transverse momenta corresponds to
$p^{\mathrm{max}}_{\mathrm{T}}=3.1$~GeV$/c$. We see from
Fig.~\ref{fig:ratios2} (right panel) that although $\delta_{W_{2}}^{2}<1$
there are  0.9~TeV points  (blue up-triangles) in the region $\tau_{\mathrm{start}}\le\tau\le
\tau_{\mathrm{end}}$ which are below 1 outside the experimental error. If we
demand that all points between $\tau_{\mathrm{start}}$ and $\tau
_{\mathrm{end}}$ should be equal to unity within experimental errors, then
$\lambda=0.22$. This is the value of $\lambda$ used in Refs.~\cite{McLerran:2014apa} and the
relevant plot is shown in the left panel of Fig.~\ref{fig:ratios2}. The
corresponding $p^{\mathrm{max}}_{\mathrm{T}}$ is shifted down to 2.9~GeV$/c$.

\begin{figure*}[t]
\centering
\includegraphics[width=6.0cm]{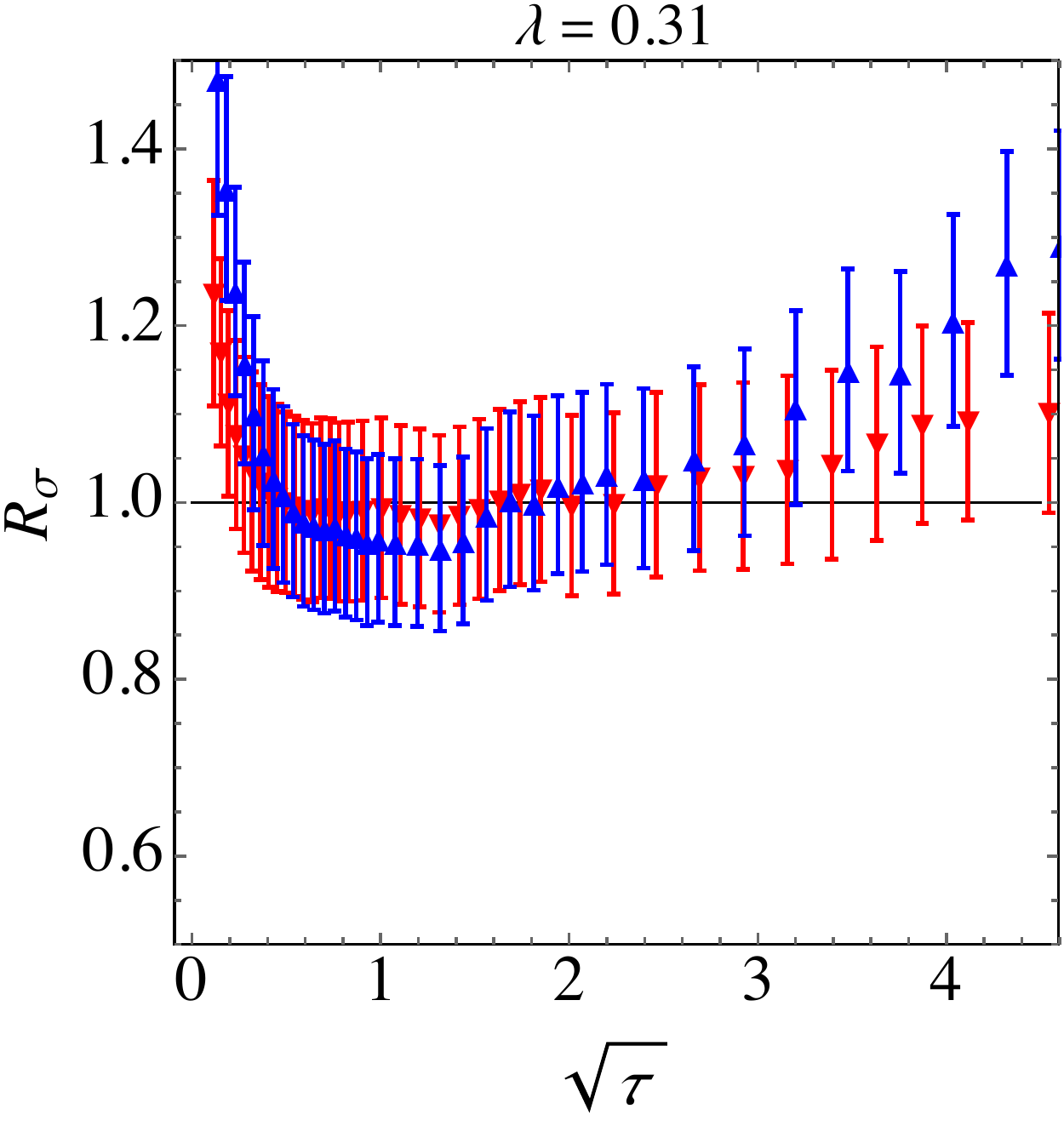}~~~
\includegraphics[width=6.0cm]{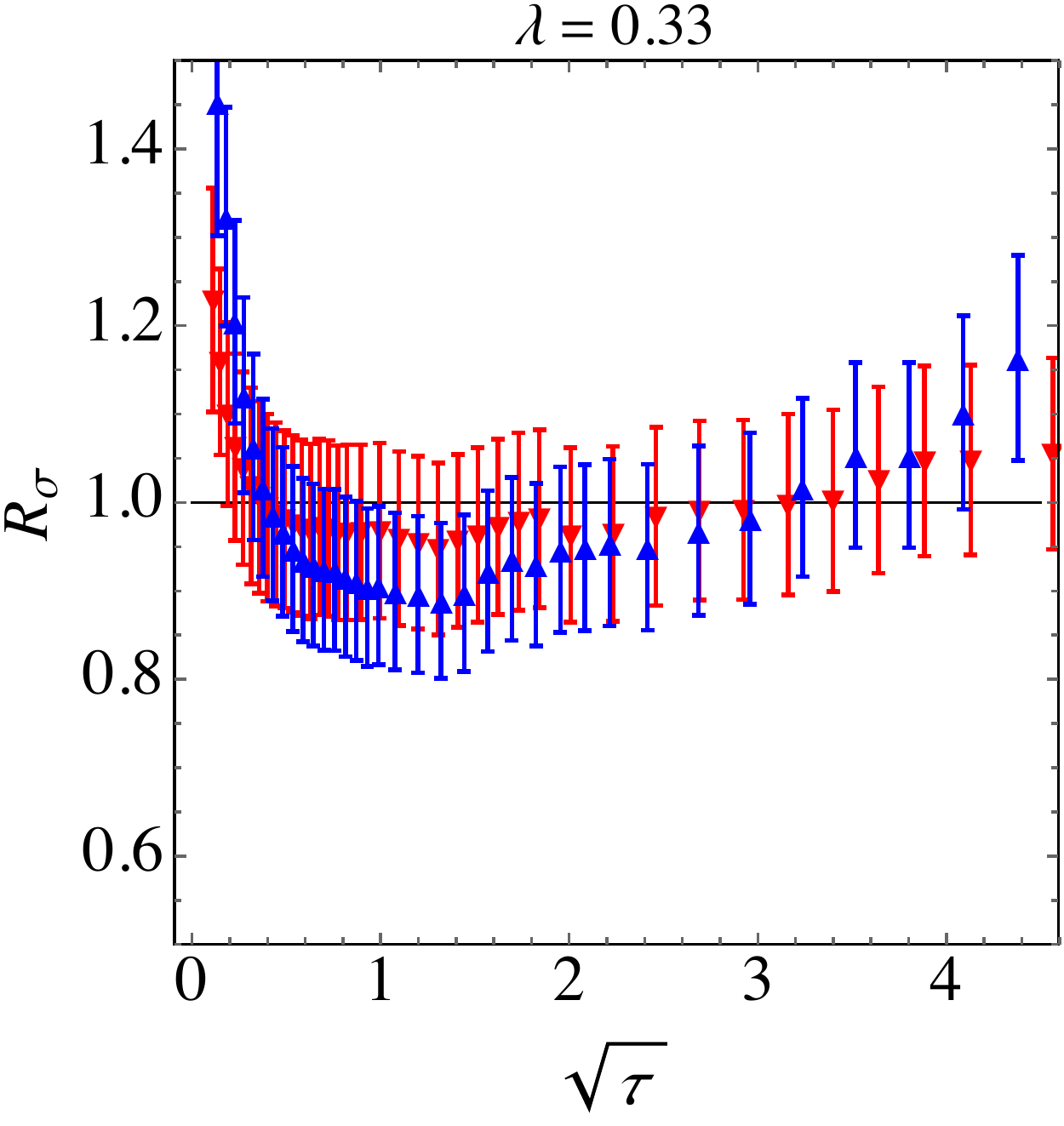}
\caption{Ratios of charged
particle cross-sections measured by ALICE, plotted as functions of  the scaling variable 
$\tau$ (\ref{tau}) for $\lambda=0.31$ (left panel) and for $\lambda=0.33$
(right panel). Red
down-triangles correspond to  to the ratio 7/2.76~TeV and blue up-triangles 
to 7/0.9~TeV. }%
\label{fig:ratios3}%
\end{figure*}

Interestingly, when we repeat this procedure for the cross-sections which are
obtained by multiplying the multiplicity spectra by the minimum bias
cross-section $\sigma_{\mathrm{MB}}(W)$ given explicitly in Ref.~\cite{Abelev:2013ala}, we
find that GS occurs at a higher value of $\lambda=0.31 - 0.33$. This by itself
is not surprising since $\sigma_{\mathrm{MB}}(W)$ depends on energy and this
dependence makes $\lambda$ different than in the case of multiplicity. What
is, however, surprising and encouraging
 is that the value of $\lambda$ is now consistent with
DIS. Moreover, the range of GS is now larger, up to $p^{\mathrm{max}%
}_{\mathrm{T}}=4.25$~GeV$/c$. This is depicted in Fig.~\ref{fig:ratios3}. The
explanation of this observation is beyond the scope of the present paper,
however, it is clear that it requires a more sophisticated model of $S_{\bot}%
$ of Eq.~(\ref{sigma_vs_N}), which in the present analysis is assumed to 
be an energy independent constant
in the case of multiplicity scaling or minimum bias cross-section in the case
of cross-section scaling.

We have performed similar analysis for the UA1 data~\cite{Albajar:1989an} 
for p$\bar{\mathrm{p}}$
cross-section at $\sqrt{s}=0.2$, 0.5 and 0.9~TeV with similar result that
$\lambda\approx0.34$. Here, however, the data extends only up to $\sim
7$~GeV$/c$ (for two lower energies) and the tail is quite noisy, namely the
ratios of the cross-sections fluctuate quite significantly for $\sqrt{\tau}
>3$.

\section{Geometrical Scaling and Tsallis parametrization}
\label{sec:GSTsallis}

It is well known that particle spectra at low and medium transverse momenta
can be described by thermal distributions in transverse mass $m_{\mathrm{T}%
}=\sqrt{p_{\mathrm{T}}^{2}+m^{2}}$ with "temperature" $T$ which is a function
of the scattering energy \cite{Hagedorn}. It is also known that more accurate
fits are obtained by means of Tsallis-like parametrization \cite{Tsallis}
where particle multiplicity distribution takes the following form (see
\emph{e.g.} \cite{Chatrchyan:2012qb}):
\begin{equation}
\frac{1}{2\pi p_{\rm  {T}}}\frac{d^{2}N_{\rm ch}}{dydp_{\rm  {T}}}=\frac{dN_{\rm ch}}{dy}%
\frac{p}{E}\frac{C_n}{2\pi}\left[  1+\frac{\widetilde{E}_{\rm  {T}}}%
{n\,T}\right]  ^{-n} \label{Tsallis}%
\end{equation}
where $\widetilde{E}_{\rm  {T}}=\sqrt{m^{2}+p_{\rm  {T}}^{2}}-m$. In what
follows we shall keep $m=0$ what implies $p/E=1$. Here%
\begin{equation}
C_n=\frac{(n-1)(n-2)}{n\,^{2}T^{2}}. \label{normC}%
\end{equation}
Coefficient $C_n$ in Eq.~(\ref{normC}) ensures proper normalization of
(\ref{Tsallis}). Indeed%
\begin{eqnarray}
\frac{dN_{\rm ch}}{dy} & = & \int\frac{1}{2\pi p_{\rm  {T}}}\frac{d^{2}N_{\rm ch}}{dydp_{\rm  {T}}%
}d^{2}p_{\rm  {T}}
\nonumber \\
& = & \frac{dN_{\rm ch}}{dy}C_n%
{\displaystyle\int\limits_{0}^{\infty}}
dp_{\rm  {T}}p_{\rm  {T}}\left[  1+\frac{p_{\rm  {T}}}{n\,T}\right]  ^{-n}
\label{dNdyint}%
\end{eqnarray}
where the last integral is equal $1/C_n.$ Here $n$ and $T$ are free fit
parameters that depend on particle species and on energy.

In the limit $n\rightarrow\infty$ (or equivalently for small $p_{\rm   T}$)
distribution (\ref{Tsallis}) tends to the
exponent
\begin{equation}
\frac{1}{p_{\rm  {T}}}\frac{d^{2}N_{\rm ch}}{dydp_{\rm  {T}}}\simeq\frac{dN_{\rm ch}}{dy}%
\frac{1}{T^{2}}\exp(-p_{\mathrm{T}}/T). \label{Thermal}%
\end{equation}
Substituting (\ref{pT}) into (\ref{Thermal}) we arraive at:%
\begin{equation}
\frac{1}{p_{\rm  {T}}}\frac{d^{2}N_{\rm ch}}{dydp_{\rm  {T}}}\simeq\frac{dN_{\rm ch}}{dy}%
\frac{1}{T^{2}(W)}\exp\left(  -\frac{\overline{Q}_{\rm  {s}}(W)}{T(W)}%
\tau^{1/(2+\lambda)}\right)  \label{tauThermal}%
\end{equation}
Equation (\ref{tauThermal}) exhibits geometrical scaling exactly, only when~\cite{Praszalowicz:2013fsa}%
\begin{equation}
T(W)=\frac{1}{\varkappa}\overline{Q}_{\rm  {s}}(W). \label{TQsat}%
\end{equation}
Then, using Eq.~(\ref{dNdy}), we get:
\begin{equation}
\frac{1}{p_{\rm  {T}}}\frac{d^{2}N_{\rm ch}}{dydp_{\rm  {T}}}\simeq\frac{b\varkappa
^{2}}{Q_{0}^{2}}\exp\left(  -\varkappa\,\tau^{1/(2+\lambda)}\right)  .
\label{exp1}%
\end{equation}
Indeed, (\ref{exp1}) is energy independent. This would generalize to the full
Tsallis distribution if exponent $n$ were constant. We know, however, from the
phenomenological fits that $n$ is decreasing with energy making $p_{\rm  {T}}$
spectra harder and -- in the same time -- introducing explicit violation of
geometrical scaling for particle spectra.

Let us observe that we can include factor $\xi$ into a definition of $b$ and
$\varkappa$, so without any loss of generality we can set $\xi=1.$ Therefore we
finally arrive at the GS-Tsallis parametrization of the $p_{\rm  {T}}$ spectra
that reads:%
\begin{eqnarray}
\left.  \frac{1}{2\pi p_{\rm  {T}}}\frac{d^{2}N_{\rm ch}}{dydp_{\rm  {T}}}\right\vert
_{W}& = & \frac{B}{Q_{0}^{2}} C_{n_W} 
\left[
1+\frac{\varkappa\,p_{\rm  {T}}}{n_{W}\,\overline{Q}_{\rm  {s}}(W)}\right]
^{-n_{W}} \label{distr}%
\end{eqnarray}
where we have introduced new constant $B=b\varkappa^{2}/2\pi$ and explicitly
indicated that $n$ is a function of $W$. This dependence would be of course a
source of GS violation. Constants $B$, $\varkappa$ and $Q_0$ should remain
energy independent.

\section{Tsallis parametrization of  ALICE data}

\label{sec:TsALICE}

In this Section we are going to check whether one can fit ALICE data~\cite{Abelev:2013ala} 
with the
help of formula (\ref{distr}).
 In the original ALICE paper~\cite{Abelev:2013ala} it is said that the multiplicity data can be fitted with
the Hagedorn distribution~\cite{Hagedorn}, rather than with the Tsallis one. 
Therefore we could
expect that ordinary fitting procedures would not give a reasonable parametrization
of the data. In order to enforce Tsallis parametrization we
have proceeded in  the following way. For each LHC energy we have chosen two
values of $p_{\rm  {T}}$, one in the small $p_{\rm  {T}}$ region and one in
the tail that are displayed in Table~\ref{tab:ptlowhigh}. For $p_{\rm  {T}}^{\rm  {low}}$
we have chosen approximately 0.5~GeV$/c$ that is already above the non-perturbative
region. For $p_{\rm  {T}}^{\rm  {high}}$ we have chosen values that are rather far from
the end of the spectrum, but already large enough to be in the perturbaitve regime.
Of course our fit parameters do depend on this choice, however, as we shall see below,
the quality of the Tsallis fits with the values of limiting $p_{\rm T}$ given in 
Table~\ref{tab:ptlowhigh} is good enough that manipulating with these values 
has  not been necessary. Let us also remark at this point that our aim here was to show
certain properties of the Tsallis fits enforced on ALICE data at low and moderate
transverse momenta,
since we new from the
beginning that this particular piece of data does not admit Tsallis parametrization in
the whole $p_{\rm T}$ range.

\begin{table}[h]
\centering
\begin{tabular}
[c]{|c|c|c|}\hline
$W\,$[TeV] & $p_{\rm  {T}}^{\rm  {low}}$ [GeV$/c$] & $p_{\rm  {T}%
}^{\rm  {high}}$ [GeV$/c$]\\\hline
0.90 & 0.525 & 8.5\\
2.76 & 0.525 & 10.5\\
7.00 & 0.525 & 13.5\\\hline
\end{tabular}
\caption{Values of $p_{\rm  {T}}^{\rm  {low}}$ and $p_{\rm  {T}}^{\rm  {high}}$
used to fit Tsallis parametrization to ALICE data (see the beginning of Sect.~\ref{sec:TsALICE}).}%
\label{tab:ptlowhigh}%
\end{table}

For the $p_{\rm  {T}}$ values given in Table~\ref{tab:ptlowhigh}
we have calculated ratios $N(p_{\rm  {T}%
}^{\rm  {low}},W)/N(p_{\rm  {T}}^{\rm  {high}},W)$, both for the data and for
parametrization (\ref{distr}). In this way normalization parameter $B$
canceled out. Now, for fixed value of $\varkappa$ we have
calculated $n_{W}$
from the following condition:%
\begin{equation}
\varkappa:\;\left.  \frac{N(p_{\rm  {T}}^{\rm  {low}},W)}{N(p_{\rm  {T}%
}^{\rm  {high}},W)}\right\vert _{\rm  {th}}=\left.  \frac{N(p_{\rm  {T}%
}^{\rm  {low}},W)}{N(p_{\rm  {T}}^{\rm  {high}},W)}\right\vert _{\rm  {exp}%
}\;\Rightarrow\;n_{W}.
\end{equation}
Note that the value of parameter $\lambda$ entering the definition of the
saturation scale (\ref{Qbarsat}) has been
already fixed by the method desribed
in Sect.~\ref{sec:GSALICE}. Here we have used $\lambda
=0.24$. 

\begin{figure}[t]
\centering
\includegraphics[height=6cm]{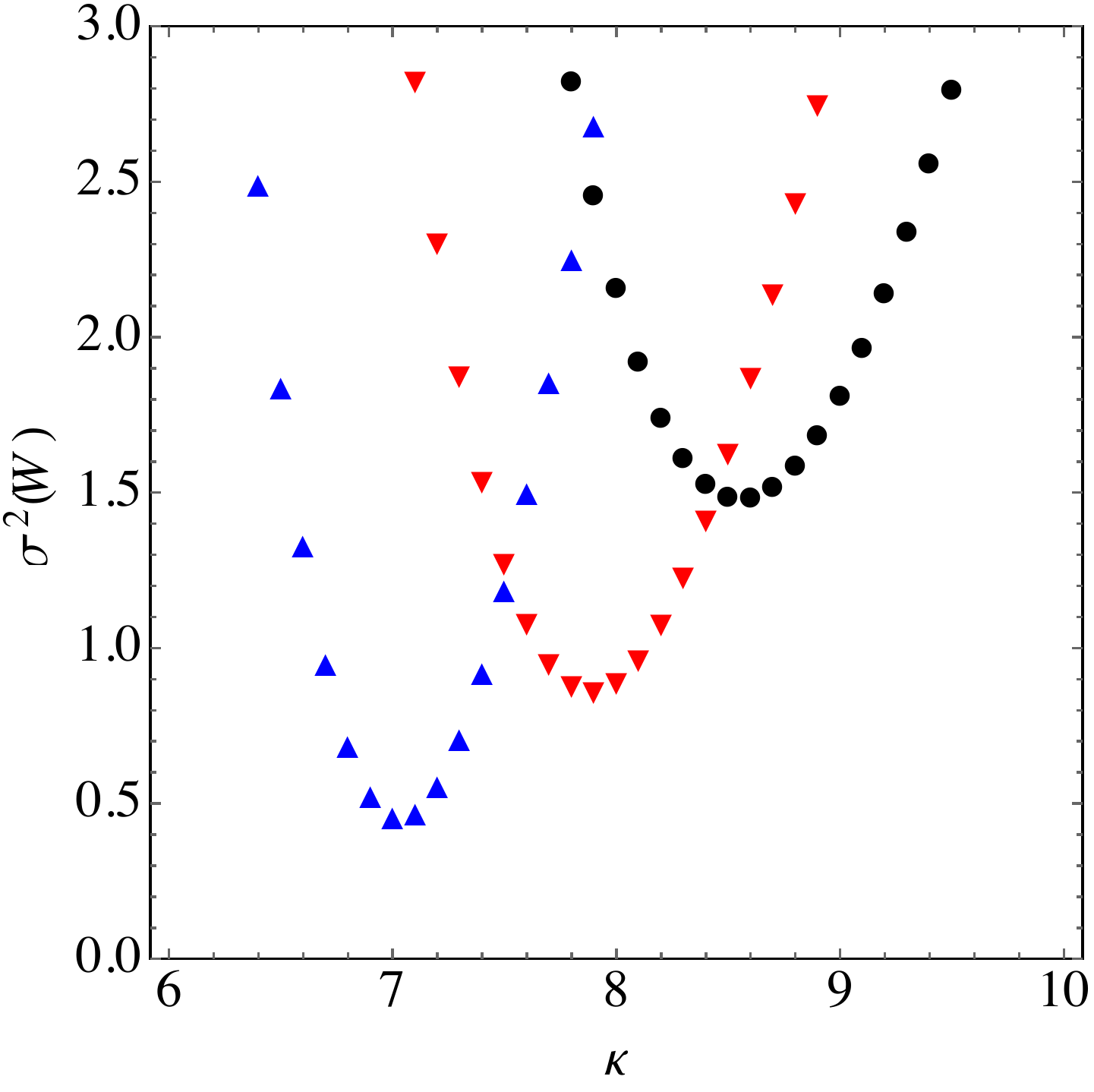}\caption{Mean square deviations
defined in Eq.(\ref{sigma}) as functions of $\varkappa$. Black circles
correspond to $W=7$~TeV, red down-triangles to 2.76~TeV and blue up-tangles to
0.9~TeV.}%
\label{fig:sigmaW}%
\end{figure}

Next, for each pair $(\varkappa,n_{W})$ we have computed mean
quadratic deviation%
\begin{equation}
\sigma_{W}^{2}(\varkappa)=\frac{1}{i^{W}_{\rm  {max}}}%
{\displaystyle\sum\limits_{i=1}^{i^{W}_{\rm  {max}}}}
\frac{\left(  \left.  N(p_{\rm  {T}}^{i},W)\right\vert _{\rm  {th}}-\left.
N(p_{\rm  {T}}^{i},W)\right\vert _{\rm  {exp}}\right)  ^{2}}{\Delta
^{2}(p_{\rm  {T}}^{i},W)} \label{sigma}%
\end{equation}
where $i$ runs over experimental data points at energy $W$ up to the maximal
$p_{\rm  {T}}$. $\Delta$ denotes the experimental error of $N$. The result is plotted
in Fig.\thinspace\ref{fig:sigmaW}. We see that functions $\sigma_{W}%
^{2}(\varkappa)$ exhibit minima at three distinct values of parameter
$\varkappa$. This is the first signal that one cannot fit ALICE data with
Tsallis distributions that correspond to the energy independent parameter
$\varkappa$. 
Therefore we have to allow for energy dependent $\varkappa
\rightarrow \varkappa_{W}$ which takes us away from the 
geometrically scaling Tsallis parametrization of Eq.(\ref{distr}).

One can also see that minima of $\sigma_{W}^{2}$ grow with
energy. This is due to the fact, that Tsallis distributions used here are not
able to describe both low $p_{\mathrm{T}}$ part and the very high end of the
spectrum simultaneously. We have checked that confining the sums in
(\ref{sigma}) to $p_{\mathrm{T}}^{\mathrm{max}} \sim20$~GeV$/c$ corresponding
to $i_{\mathrm{max}}=54$ for all energies in question, brings down $\sigma
_{W}^{2}$ below 0.8 (see Table~\ref{tab:allpars}). This means that Tsallis
parameterization with energy dependent  $\varkappa_W$ 
used here is able to describe the $p_{\mathrm{T}}$ spectra
at small and moderate transverse momenta, \emph{i.e.} precisely in the region
we are interested in.

In Table~\ref{tab:allpars} we collect values of the parameters $\varkappa_W$,
$n_{W}$ and $B_{W}$ at the minima of $\sigma_{W}^{2}$. $B_{W}$ is calculated
from the condition $\left.  N(p_{\rm  {T}}^{\rm  {low}},W)\right\vert
_{\rm  {th}}=\left.  N(p_{\rm  {T}}^{\rm  {low}},W)\right\vert _{\rm  {exp}}$.
The resulting spectra together with ALICE data are shown in
Fig.~\ref{fig:spectra} both for distributions expressed in terms of $p_{\rm   T}$
and in terms of scaling variable $\sqrt{\tau}$. The quality of this fit can be also 
appreciated by looking at the right panel of Fig.~\ref{fig:ratios2} where the 
multiplicity ratios are well reproduced without any adjustment of fit parameters.

\begin{table}[ptb]
\centering
\begin{tabular}
[c]{|c|c|c|c|c|c|}\hline
$W\,$[TeV] & $\varkappa_{W}$ & $n_{W}$ & $B_{W}$ & \multicolumn{2}{c|}{$\sigma
_{W}^{2}$}\\
&  &  &  & all $p_{\rm  {T}}$ & $p_{\rm  {T}}<20$ GeV$/c$\\\hline
0.90 & 7.0 & 8.32 & 28.88 & 0.46 & 0.46\\
2.76 & 7.9 & 7.33 & 36.96 & 0.87 & 0.71\\
7.00 & 8.6 & 6.79 & 43.82 & 1.49 & 0.75\\\hline
\end{tabular}
\caption{Parameters entering Tsallis parametrization (\ref{distr}) coming from the fit
to  ALICE spectra and the corresponding values of mean square deviation for all
measured $p_{\rm   T}$ values and for $p_{\rm  {T}}<20$ GeV$/c$.}%
\label{tab:allpars}%
\end{table}

Given the fact that the saturation momentum scales as a power of energy
$\overline{Q}_{\mathrm{s}}(W)=(W/Q_{0})^{\lambda_{\mathrm{eff}}}$ we have
tried to fit energy dependence of parameters $\kappa_{W}$, $n_{W}$ and $B_{W}$
with generic form $a_{0}(W/Q_{0})^{\alpha}$ with the following result:
\begin{align}
\kappa_{W}  &  =7.097\;(W/Q_{0})^{0.1000},\nonumber\\
n_{W}  &  =8.199\;(W/Q_{0})^{-0.1005},\nonumber\\
B_{W}  &  =29.76\;(W/Q_{0})^{0.2013}. \label{fits}%
\end{align}
This result is surprisingly in line with the effective exponent of the
saturation scale which for $\lambda=0.24$ reads $\lambda_{\mathrm{eff}%
}=0.1071$. Note also that $B\sim\kappa^{2}$ (see definition of $B$ below
Eq.~(\ref{distr})), and this dependence is reproduced by the fits of
Eqs.~(\ref{fits}). Although this energy dependence follows only from the fit
to data, and we do not have any model to explain their values, it is a
reasonable assumption to take:
\begin{align}
\kappa_{W}  &  =\kappa_{0}\;\frac{\overline{Q}_{\mathrm{s}}(W)}{Q_{0}%
},\nonumber\\
n_{W}  &  =n_{0}\;\frac{Q_{0}}{\overline{Q}_{\mathrm{s}}(W)},\nonumber\\
B_{W}  &  =B_{0}\;\frac{\overline{Q}_{\mathrm{s}}^{2}(W)}{Q_{0}^{2}}
\label{fitsQsat}%
\end{align}
where the coefficients $\kappa_{0}$, $n_{0}$ and $B_{0}$ can be read off from
Eq.~(\ref{fits}). In what follows we shall drop $Q_{0}=1$~GeV/$c$ which was
included in (\ref{fitsQsat}). Therefore we have:%
\begin{align}
\left.  
\frac{d^{2}N_{\rm ch}}{dyd^2p_{\rm  {T}}}\right\vert
_{W}  &  =B_{0}\,\overline{Q}_{\mathrm{s}}^{2}(W)\,\frac{(n_{0}-\overline
{Q}_{\mathrm{s}}(W))(n_{0}-2\overline{Q}_{\mathrm{s}}(W))}{n_{0}^{2}%
}\nonumber\\
&  \times\left[  1+\frac{\varkappa_{0}}{n_{0}%
}\overline{Q}_{\rm  {s}}^{2}(W)\tau^{1/(2+\lambda)}\right]  ^{-n_{0}%
/\overline{Q}_{\mathrm{s}}(W)}. \label{NchTsallis}%
\end{align}

For geometrical scaling to be present we need this function to be inedependent
of $W$, \emph{i.e.} independent of $\overline{Q}_{\mathrm{s}}(W).$ For the
energies in question (from a few hundreds GeV up to a few TeV) $\overline
{Q}_{\mathrm{s}}(W)$ changes from 0.9 to 1.5. Therefore the factor involving
$n_{0}$ is in fact close to 1 and the main energy dependence comes from
$\overline{Q}_{\mathrm{s}}^{2}(W)$ in front 
and from the factor in a square bracket in
Eq.~(\ref{NchTsallis}). 

\begin{figure*}[t]
\centering
\includegraphics[height=5.8cm]{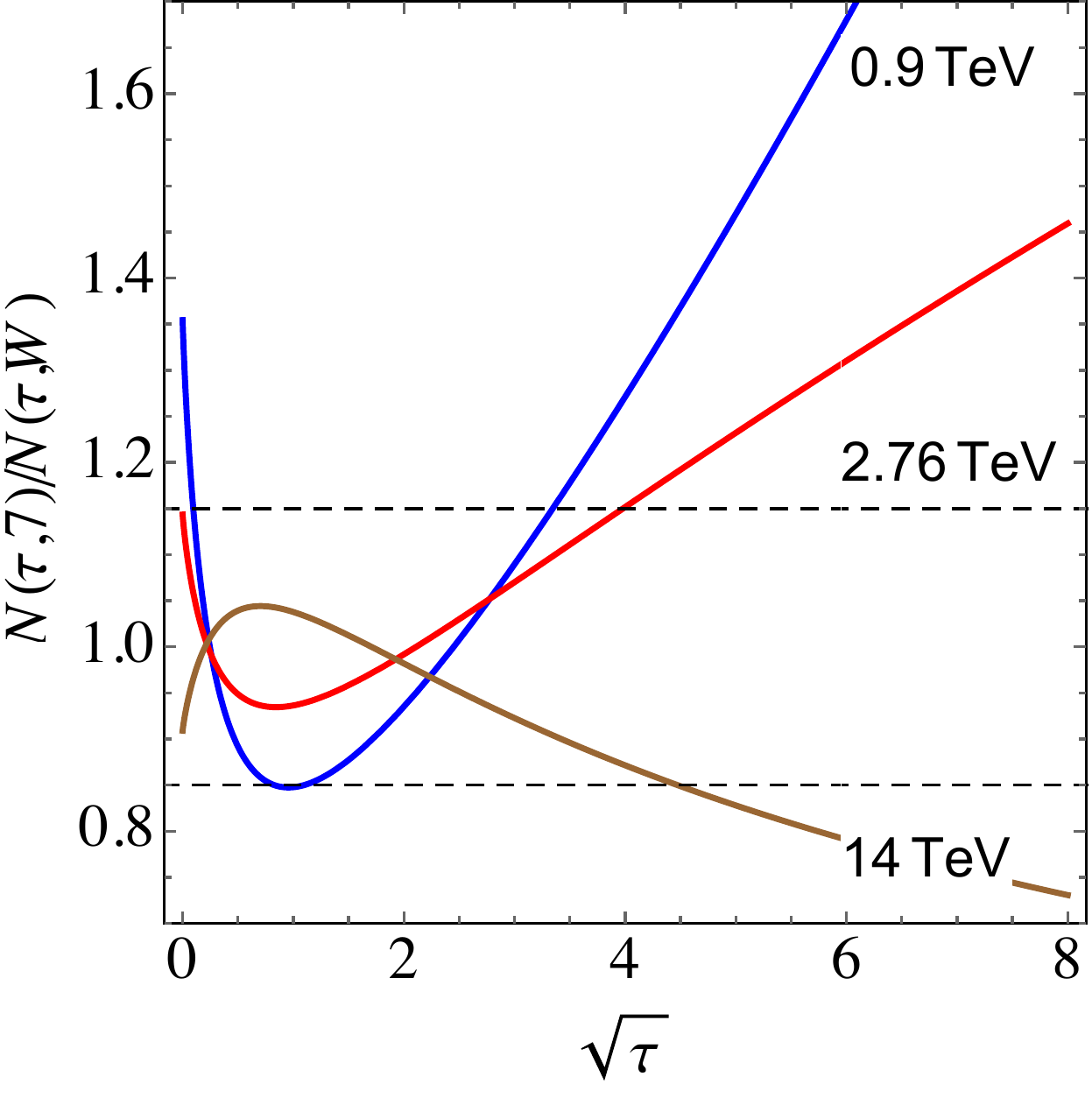}~~
\includegraphics[height=5.9cm]{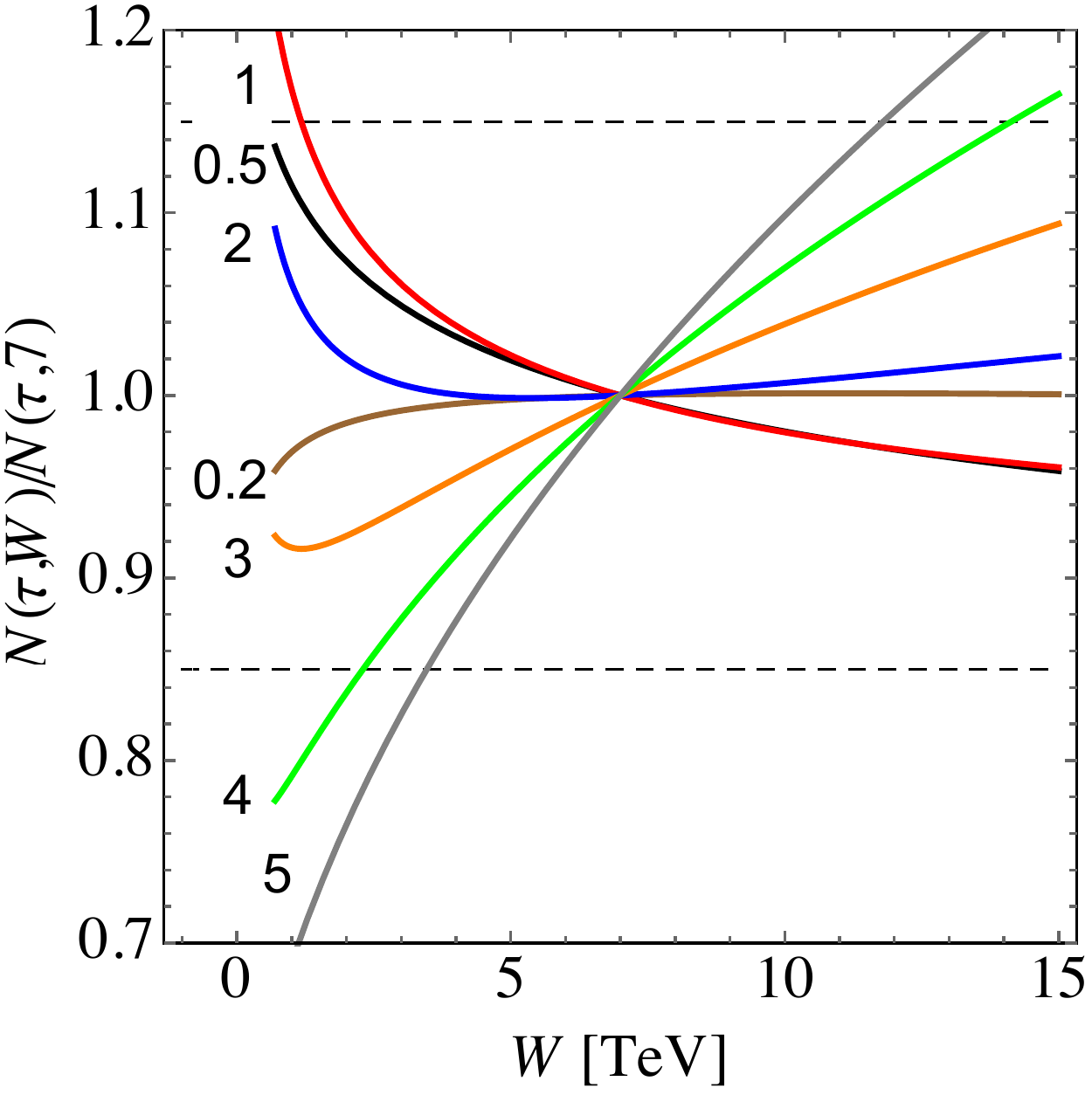}
\caption{Left panel: Ratios (\ref{Rdef}) in terms of parametrization of Eq.~(\ref{NchTsallis})
for $W'=0.9$ (blue), 2.76 (red)  and 14~TeV (brown). Right panel: multiplicities 
$N(\sqrt{\tau_i},W)$ normalized to $N$ at 7~TeV plotted as functions of $W$ in TeV for
fixed values of $\sqrt{\tau_i}$ displayed next to the vertical axis.}%
\label{fig:NbyN}%
\end{figure*}

In order to see how GS is reached by
Eq.~(\ref{NchTsallis}) we plot in Fig.~\ref{fig:NbyN} ratio $N(\stau
,7)/N(\stau,W)$ as a function of $\sqrt{\tau}$ for $W=0.9$, 2.76 and 14~TeV
(left panel). Horizontal dashed lines at $1\pm0.15$ show 15\% band around unity
which roughly corresponds to the size of the experimental errors $\Delta_R$ (10 \%)
and the accuracy of the fit (5 \%) -- see Fig.~\ref{fig:ratios}. GS is present if theoretical solid curves
fall within this interval.
We can conclude from Fig.~\ref{fig:NbyN} that with this accuracy GS should be seen
in the data up to $\sqrt{\tau} \sim 4$ for the whole LHC energy range up to 14~TeV.
Should Tsallis parametrization (\ref{NchTsallis}) hold for higher energies we might
expect shrinking of the maximal $\sqrt{\tau}$ where GS is still present with increasing energy.
Given the fact that for fixed $\tau$ transverse momentum $p_{\rm   T}$ is an increasing function 
of energy
(see Eq.~(\ref{pT})), this may not immediately mean that the $p_{\rm   T}$ window
for GS would be shrinking as well.

The same conclusion can be reached by looking at Eq.~(\ref{NchTsallis})
where $N(\stau,W)/N(\stau,7)$ is plotted
as a function of energy for fixed $\tau$.
In the right panel of Fig.~\ref{fig:NbyN}
we plot $N(\stau,W)/N(\stau,7)$ as a function
of  $W$ for different values of $\sqrt{\tau}=0.2$, 0.5, 1, 2, 3, 4 and
5 shown next to the horizontal axis.

\section{Conclusions}
\label{sec:Concl}

In this paper we have addressed three questions concerning saturation in high energy
pp scattering. To this end we have used recent ALICE data on inelastic scattering at the
LHC \cite{Abelev:2013ala}. 

The first question concerned the very existence of geometrical scaling in
multiplicity distributions. By applying a model-independent {\em method of ratios}
we have established that GS is indeed present in multiplicity spectra over a limited
transverse momentum range up to $\sim 3$~GeV$/c$ with characteristic exponent
$\lambda \sim 0.22 - 0.24$. This exponent is significantly different than in DIS,
where $\lambda=0.32$,
 and
also lower than the one extracted from the CMS non-single diffractive data: $\lambda=0.27$.
We have proposed the solution to this discrepancy by looking at GS for the inelastic
cross-section rather than for the multiplicity distribution. Motivation for this comes from the
$k_{\rm   T}$ factorized form of the gluon production in pp collisions (\ref{Nchdef}) that
leads straightforwardly to Eq.~(\ref{sigmaGS})
and from the fact that the proportionality factor between the multiplicity and the cross-section
(\ref{sigma_vs_N}) is not energy-independent. We have found that the inelastic cross-section
scales better than multiplicity up to he maximal transverse momentum that is larger than 4~GeV$/c$
and with the characteristic exponent $\lambda \sim 0.31 - 0.33$. We have also looked at the UA1 data
for $\bar{\rm   p}$p scattering and obtained similar value of $\lambda$. We believe that this
observation provides a solution to the discrepancy between scaling properties in DIS and 
in hadronic collisions.

The second question concerned the universal shape of GS and its connection to the Tsallis
distribution. We have confirmed that the natural answer to this question is provided by
a parametrization where the Tsallis "temperature" $T$ is proportional to the average
saturation scale $\overline{Q}_{\rm   s}$ (\ref{Qbarsat}) and the remaining 
Tsallis parameter $n$ should be an 
energy independent constant. In practice $n$ does depend on energy and this leads to
the violation of GS for this particular parametrization.

Finally the third question was whether such a simple solution is admitted by the experimental
data. We have found that recent ALICE data on inelastic charged particle multiplicity
does not admit the above solution, in agreement with the original claim of  
Ref.~\cite{Abelev:2013ala}. We have found another parametrization where Tsallis
parameter $n$ is inversely proportional to $\overline{Q}_{\rm   s}$. This parametrization
indeed exhibits GS in the limited energy range, however GS is not obviously extended
to higher energies. We have concluded at this point that the solution we found was rather
accidental. It will be therefore interesting to see whether this solution will be still present at
higher energies of the LHC run II.

\section*{Acknowledgements} 
This research  has been financed in part by
the Polish NCN grant 2014/13/B/ST2/02486.


\begin{thebibliography}{99} 

\bibitem {Mueller:2001fv}A.~H.~Mueller, 
arXiv:hep-ph/0111244.


\bibitem {McLerran:2010ub}L.~McLerran, 
{Acta
Phys.\ Pol.\ B} {\bf 41} (2010) 2799 [arXiv:1011.3203 [hep-ph]].

\bibitem {BK}I.~Balitsky,
Nucl.\ Phys.\ B {\bf 463} (1996) 99; \\
Y.~V.~Kovchegov,
Phys.\ Rev.\ D {\bf 60} (1999) 034008 and
Phys.\ Rev.\  D {\bf 61} (2000) 074018.


\bibitem {jimwlk}J. Jalilian-Marian, A. Kovner, A. Leonidov, and H. Weigert,
Nucl. Phys. B {\bf 504} (1997)  415  and Phys. Rev. D {\bf 59}
(1998) 014014; \\ 
E. Iancu, A. Leonidov, and L. D. McLerran, Nucl. Phys. A  {\bf 692},
(2001) 583 ; \\
E. Ferreiro, E. Iancu, A. Leonidov, and L. D. McLerran, Nucl.
Phys. A  {\bf 703} (2002) 489.

\bibitem {Munier:2003vc} S.~Munier, and R.~B.~Peschanski,
Phys.\ Rev.\ Lett.\ {\bf 91} (2003) 232001
[hep-ph/0309177] and
Phys.\ Rev.\ D {\bf 69} (2004) 034008
[hep-ph/0310357].

\bibitem{Stasto:2000er}
  A.~M.~Stasto, K.~J.~Golec-Biernat and J.~Kwiecinski,
  Phys.\ Rev.\ Lett.\  {\bf 86} (2001) 596
  [hep-ph/0007192].

\bibitem {MLV} L.~D.~McLerran, and R.~Venugopalan,
Phys.\ Rev.\ D  {\bf 49} (1994) 2233 , \ Phys. Rev.  D {\bf 49}
(1994) 3352  \  and Phys.\ Rev.\ D {\bf  50} (1994) 2225 .

\bibitem{fn1}
Throughout
this paper we shall generically denote by $\cal F$ an energy independent function of scaling
variable $\tau$, although in some cases some constant factors are included in   $\cal F$
without changing notation.

\bibitem{McLerran:2010ex}
  L.~McLerran and M.~Praszalowicz,
  Acta Phys.\ Polon.\ B {\bf 41} (2010) 1917
  [arXiv:1006.4293 [hep-ph]] and
  Acta Phys.\ Polon.\ B {\bf 42} (2011) 99
  [arXiv:1011.3403 [hep-ph]].
  
\bibitem{Praszalowicz:2012zh}
  M.~Praszalowicz and T.~Stebel,
  JHEP {\bf 1303} (2013) 090
  [arXiv:1211.5305 [hep-ph]] and
  JHEP {\bf 1304} (2013) 169
  [arXiv:1302.4227 [hep-ph]].
  
\bibitem{Praszalowicz:2013uu}
  M.~Praszalowicz,
  Phys.\ Rev.\ D {\bf 87} (2013)  071502
  [arXiv:1301.4647 [hep-ph]].
  
\bibitem{Abelev:2013ala}
  B.~B.~Abelev {\it et al.}  [ALICE Collaboration],
  Eur.\ Phys.\ J.\ C {\bf 73} (2013) 12,  2662
  [arXiv:1307.1093 [nucl-ex]].
  
\bibitem{Tsallis}
C. Tsallis, J. Stat. Phys. {\bf 52} (1988) 479 and Eur. Phys. J. A {\bf 40} (2009) 257;\\
T. S. Bir{\'o}, G. Purcsel,
and K. {\"U}rm{\"o}ssy,
Eur. Phys. J. \rm  bf{A 40}, 325 (2009).


\bibitem{Wong:2012zr}
  C.~Y.~Wong and G.~Wilk,
  Acta Phys.\ Polon.\ B {\bf 43} (2012) 2047
  [arXiv:1210.3661 [hep-ph]] and
  Phys.\ Rev.\ D {\bf 87} (2013) 11,  114007
  [arXiv:1305.2627 [hep-ph]] and
  arXiv:1309.7330 [hep-ph].
  
 \bibitem{Cirto:2014sra}
  L.~J.~L.~Cirto, C.~Tsallis, C.~Y.~Wong and G.~Wilk,
  arXiv:1409.3278 [hep-ph];\\
  C.~Y.~Wong, G.~Wilk, L.~J.~L.~Cirto and C.~Tsallis,
  EPJ Web Conf.\  {\bf 90} (2015) 04002
  [arXiv:1412.0474 [hep-ph]].
  
\bibitem{Cleymans:2013rfq}
  J.~Cleymans, G.~I.~Lykasov, A.~S.~Parvan, A.~S.~Sorin, O.~V.~Teryaev and D.~Worku,
  Phys.\ Lett.\ B {\bf 723} (2013) 351
  [arXiv:1302.1970 [hep-ph]];\\
  M.~D.~Azmi and J.~Cleymans,
  J.\ Phys.\ G {\bf 41} (2014) 065001
  [arXiv:1401.4835 [hep-ph]];\\
   L.~Marques, J.~Cleymans and A.~Deppman,
  Phys.\ Rev.\ D {\bf 91} (2015) 054025
  [arXiv:1501.00953 [hep-ph]].
  
\bibitem{Gribov:1981kg}
  L.~V.~Gribov, E.~M.~Levin and M.~G.~Ryskin,
  Phys.\ Lett.\ B {\bf 100} (1981) 173.
  
\bibitem{Szczurek:2003vq}
  A.~Szczurek,
  Acta Phys.\ Polon.\ B {\bf 35} (2004) 161
  [hep-ph/0311175].
  
 \bibitem {GolecBiernat:1998js} K.J.~Golec-Biernat and M.~W{\"u}sthoff,
Phys.\ Rev.\ D \rm  bf{59} (1998) 014017  [hep-ph/9807513] and
Phys.\ Rev.\ D \rm  bf{60} (1999) 114023  [hep-ph/9903358].


  
\bibitem{Kharzeev:2002ei}
  D.~Kharzeev, E.~Levin and M.~Nardi,
  Nucl.\ Phys.\ A {\bf 730} (2004) 448
  [Nucl.\ Phys.\ A {\bf 743} (2004) 329]
  [hep-ph/0212316] and
  Nucl.\ Phys.\ A {\bf 747} (2005) 609
  [hep-ph/0408050].
\bibitem{Kowalski:2003hm}
  H.~Kowalski and D.~Teaney,
  Phys.\ Rev.\ D {\bf 68} (2003) 114005
  [hep-ph/0304189].
  
\bibitem{Tribedy:2010ab}
  P.~Tribedy and R.~Venugopalan,
  Nucl.\ Phys.\ A {\bf 850} (2011) 136
   [Nucl.\ Phys.\ A {\bf 859} (2011) 185]
  [arXiv:1011.1895 [hep-ph]].
  
\bibitem{Levin:2011hr}
  E.~Levin and A.~H.~Rezaeian,
  Phys.\ Rev.\ D {\bf 83} (2011) 114001
  [arXiv:1102.2385 [hep-ph]].
\bibitem{PHD}
Ya.I. Azimov, Yu.L. Dokshitzer, V.A. Khoze and S.I. Troian, Z. Phys. C {\bf 27} (1985) 65;\\
Yu. L. Dokshitzer, V. A. Khoze and S. I. Troyan, J. Phys. G
{\bf 17} (1991) 1585;\\
V.A. Khoze and W. Ochs, Int. J. Mod. Phys. A {\bf 12} (1997) 2949;\\
S. Lupia and W. Ochs, Phys. Lett. B {\bf 418} (1998) 214.

\bibitem{McLerran:2014apa}
  L.~McLerran and M.~Praszalowicz,
  Phys.\ Lett.\ B {\bf 741} (2015) 246
  [arXiv:1407.6687 [hep-ph]];\\
   M.~Praszalowicz,
  AIP Conf.\ Proc.\  {\bf 1654} (2015) 080001
  [arXiv:1410.5220 [hep-ph]].
  
\bibitem{Albajar:1989an}
  C.~Albajar {\it et al.}  [UA1 Collaboration],
  Nucl.\ Phys.\ B {\bf 335} (1990) 261.
 
 \bibitem {Hagedorn}R. Hagedorn, Nuovo Cim. Suppl. {\bf 3} (1965) 147.
 
 \bibitem {Chatrchyan:2012qb} S.~Chatrchyan \rm  it{et al.} [CMS
Collaboration],
Eur.\ Phys.\ J.\ C {\bf 72}, 2164 (2012)  [arXiv:1207.4724 [hep-ex]].

\bibitem{Praszalowicz:2013fsa}
  M.~Praszalowicz,
  Phys.\ Lett.\ B {\bf 727} (2013) 461
  [arXiv:1308.5911 [hep-ph]].
  
\end{thebibliography}
\end{document}